# Bragg spectroscopy of a superfluid Bose-Hubbard gas

X. Du, Shoupu Wan, Emek Yesilada, C. Ryu, and D. J. Heinzen

*Department of Physics, The University of Texas at Austin, Austin, TX 78712*

Z. X. Liang[1,2] and Biao Wu[1]

[1]*Institute of Physics, Chinese Academy of Sciences, P. O. Box 603, Beijing 100080, China*

[2]*Institute of Metal Research, Chinese Academy of Sciences, Shenyang 110016, China*

Bragg spectroscopy is used to measure excitations of a trapped, quantum-degenerate gas of $^{87}$Rb atoms in a 3-dimensional optical lattice. The measurements are carried out over a range of optical lattice depths in the superfluid phase of the Bose-Hubbard model. For fixed wavevector, the resonant frequency of the excitation is found to decrease with increasing lattice depth. A numerical calculation of the resonant frequencies based on Bogoliubov theory shows a less steep rate of decrease than the measurements.

PACS numbers: 03.75.Kk, 03.75.Lm, 05.30.Jp, 32.80.Pj

Quantum-degenerate atoms in optical lattices form a strongly interacting many-body system whose parameters can be readily controlled. As first pointed out by Jaksch *et al.*, [1] bosonic atoms in an optical lattice constitute a nearly ideal realization of the Bose-Hubbard model [2]. This model predicts a superfluid to Mott insulator quantum phase transition that has been observed by Greiner *et al.* [3]. Since then, this field has attracted great interest due to its potential for the realization of quantum computation and quantum simulation of strongly-correlated many-body systems [4].

A key property of a quantum gas is its excitation spectrum. Excitations of a Bose-Hubbard gas by a gradient of magnetic field [3] or a modulated optical lattice depth [5, 6] have been previously observed. However, neither of these techniques directly probes the linear excitation spectrum of the gas, since a tilted lattice perturbs the gas only at zero frequency, and a modulated optical lattice only at zero quasi-momentum. The latter case results in a nonlinear excitation spectrum that has been analyzed only very recently [7]. Bragg spectroscopy has been demonstrated as a probe of the linear excitation spectrum of a Bose-Einstein condensate [8-11], and has also been proposed as a method to study the Mott insulator phase of the Bose-Hubbard gas [12, 13]. In this paper, we report the first application of Bragg spectroscopy to a quantum-degenerate Bose gas in a 3-D optical lattice. We observe resonant excitation of the gas in the superfluid regime of the Bose-Hubbard model, and find that the frequency decreases with increasing lattice depth. We carry out a numerical calculation of the resonant frequencies based on Bogoliubov theory, and find that the calculated frequencies decrease at a somewhat lower rate than the measured ones.

In our experiment, a Bose-Einstein condensate (BEC) of about N = $5\times10^5$ $^{87}$Rb atoms in the hyperfine state $|F=1, m_F=-1\rangle$ is prepared in a "cloverleaf" magnetic trap [14], with an axial trapping frequency of 11.6 Hz and radial trapping frequency of 20.7 Hz. The condensate has an ellipsoidal shape and an inverted parabolic density profile n(**r**) [15] with Thomas-Fermi axial and radial radii of about $z_0 = 26\ \mu m$ and $r_0 = 15\ \mu m$, respectively. The initial condensate fraction is greater than 90%, and its peak density is $n_0 = (5.2\pm1)\times10^{13}\ cm^{-3}$.

A 3-D optical lattice is created with three mutually orthogonal optical standing waves, formed by three retroreflected, linearly-polarized laser beams from a single-frequency Ti:Sapphire laser with a wavelength $\lambda_L$ = 830 nm. The "axial" lattice laser beams propagate parallel to the symmetry axis of the BEC, and have a spot size (1/$e^2$ intensity radius) at the BEC of 130 $\mu m$. The two sets of "radial" lattice beams have a spot size at the BEC of 260 $\mu m$. Frequency shifts of several tens of MHz between these three beam pairs suppress the effects of interference between them. The resulting optical dipole potential has the form $V(x,y,z) = V_0 \left( \sin^2(k_L x) + \sin^2(k_L y) + \sin^2(k_L z) \right)$, with $k_L = 2\pi/\lambda_L$, and lattice constant $a = \lambda_L/2 = 0.415\ \mu m$. The lattice height $V_0$ is calibrated to an accuracy of about 10% with measurements of the diffraction pattern of the atoms from a short pulse of each beam pair. This gas provides a realization of the Bose-Hubbard model [1], and for our parameters the gas is entirely superfluid below the critical lattice height $V_{0c} \approx 13\ E_R$, where $E_R = \hbar^2 k_L^2 / 2m = h \times 3.33\ kHz$ is the lattice recoil energy, and m the atomic mass.

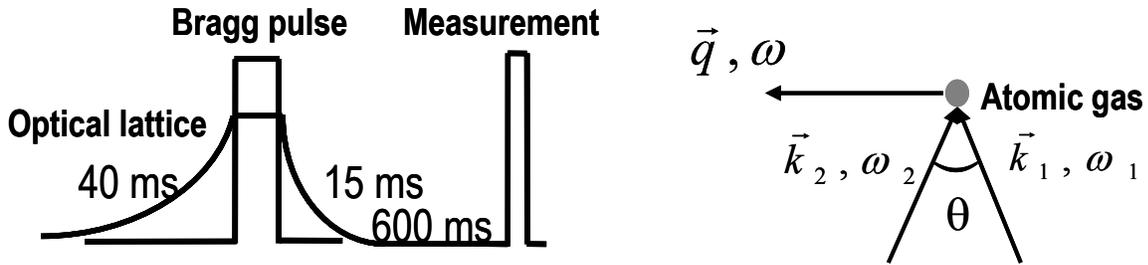

Fig. 1. (a) Time sequence of the experiment. (b) Schematic diagram of Bragg spectroscopy.

The experimental sequence is illustrated in Fig. 1(a). The BEC is loaded into the optical lattice by ramping up the lattice laser intensity with a quadratic function in a time period of 40 ms. After the lattice is turned on, we induce Bragg excitation with two laser beams of wavevector $\mathbf{k}_1$ and $\mathbf{k}_2$, and frequencies $\omega_1$ and $\omega_2$, respectively, as illustrated in Fig. 1(b). These two beams perturb the gas with a traveling wave optical dipole potential $V_B \cos(\mathbf{q} \cdot \mathbf{r} - \omega t)$ with wavevector $\mathbf{q} = \mathbf{k}_1 - \mathbf{k}_2$ and frequency $\omega = \omega_1 - \omega_2$. The Bragg beams are 400 GHz red-detuned from the atomic transition $5s_{1/2} \to 5p_{3/2}$, and have intensities between 70 and 320 mW/cm$^2$, corresponding to $V_B$ in the range from 0.45 to 1.3 $E_R$. In our experiment, the Bragg wavevector **q** is held fixed and directed perpendicular to the symmetry axis of the gas and at an angle of $\pm 45°$ with respect to the radial lattice beams. Bragg pulse durations are 20 ms for zero and 1.1 $E_R$ lattice depth, 5 ms for 2.2 $E_R$ lattice depth, and vary from 2 to 5 ms for higher lattice depths. The angle between the two Bragg beams is $30.6° \pm 0.6°$, corresponding to $q = 4.25 \pm 0.08$ $\mu m^{-1}$. Thus, the Bragg wavelength is $\lambda_B = 2\pi/q = 1.48$ $\mu m$, which corresponds to 3.56 lattice constants.

The Bragg pulse produces excitations in the gas with wavevector **q**. In order to measure the degree of excitation of the gas, we ramp the lattice beams back down to zero depth in 15 ms, wait for 600 ms for the gas to rethermalize, and finally measure the condensate fraction with time-of-flight absorption imaging. Excitation of the gas leads to an increased final temperature of the gas after thermalization, or equivalently to a reduced condensate fraction. We used this procedure rather than measuring outcoupled atoms [8-11] because at the higher lattice strengths, the gas acquires a large momentum spread due to quantum depletion [3, 16], and this makes it difficult to cleanly observe the Bragg-diffracted atoms in a time-of-flight image.

Fig. 2 shows experimental Bragg spectra recorded at five different lattice depths ranging from 0 to 9.9 $E_R$. In each case, a resonant heating of the gas is observed. The results differ substantially from previous zero quasi-momentum excitation studies [5,6], where almost no excitations were observed for lower optical lattice depths and a broad resonance appeared for higher optical lattice depths. We have searched for and found no dependence of the resonance frequencies on the Bragg beam intensities and pulse durations that is significant relative to the experimental error. The heating which appears off-resonance for lattice strengths greater than about 7 $E_R$ is due to non-adiabatic effects rather than to Bragg excitation.

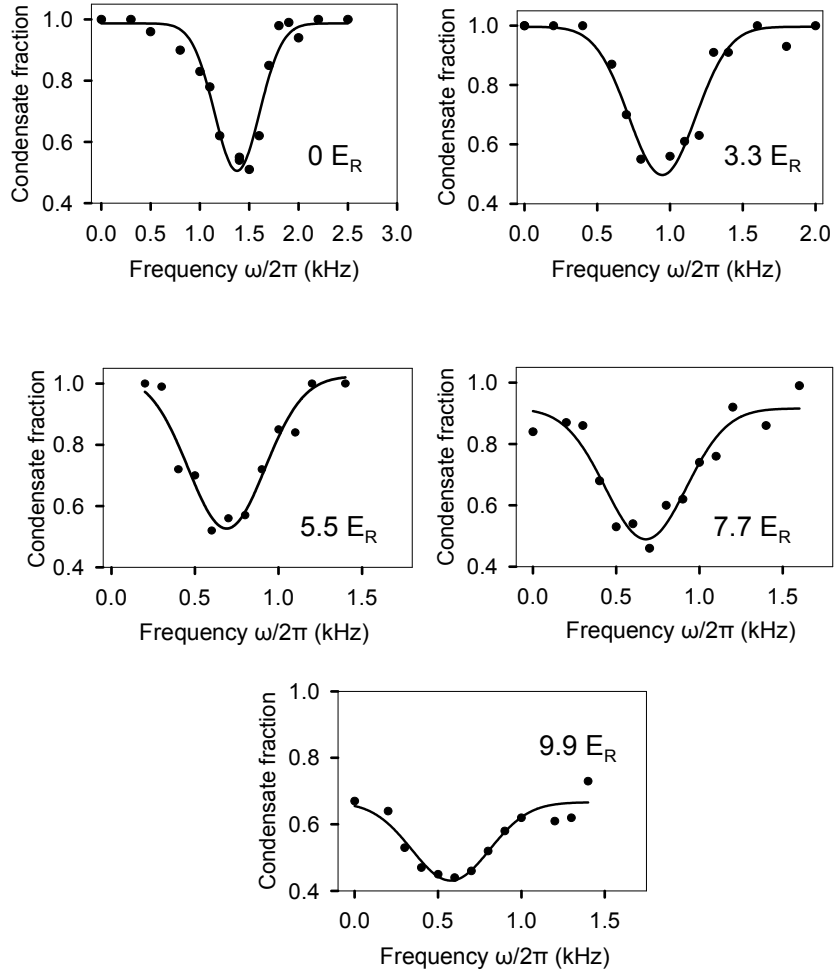

Fig. 2. Bragg spectra at optical lattice depths $V_0 = 0$, 3.3, 5.5, 7.7, and 9.9 $E_R$. Each peak is fitted with a Gaussian.

The results of the experiment are summarized in Fig. 3. Approximately 40 spectra total were taken for ten different lattice depths, and from each spectrum we obtained a resonance frequency $\omega_0$ and an rms width $\Delta\omega$ from a Gaussian fit to the data. Fig. 3 shows the measured resonance frequency as a function of lattice depth. Each data point is an average over several measurements, and the error bars indicate the statistical variation between measurements.

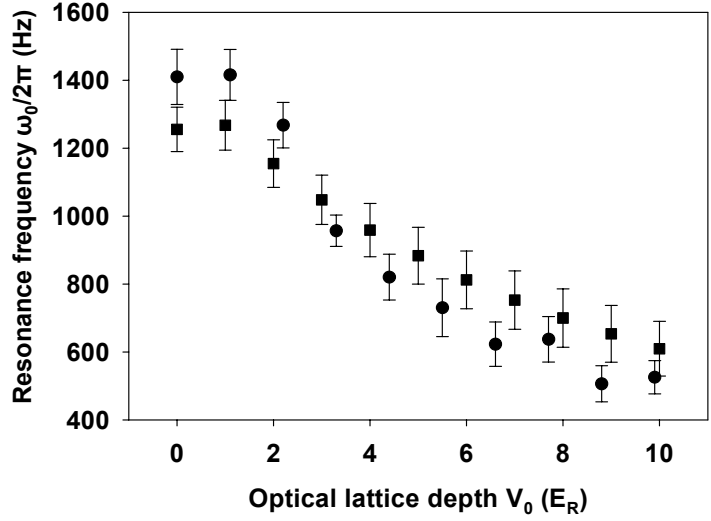

Fig. 3. Comparison between the experimental (circles) and theoretical values (squares) of the resonant excitation frequency versus optical lattice depth. The theory shows the calculated frequency for the density n = 0.57 n$_0$(V$_0$).

At zero lattice strength, the results correspond to previous measurements of Bragg excitation in condensates. In our experiment, the Bragg wavelength is small relative to the condensate radius $r_0$, and the Doppler spread $\hbar q/mr_0$ of the Bragg resonance is much less than $\mu/\hbar$, where $\mu = gn_0 = h \times 404\,Hz$ is the chemical potential of the gas, with $g = 4\pi\hbar^2 a/m$, and $a = 5.31\,nm$ [17] the scattering length. Therefore a local density approximation applies [8-11], and the gas is resonantly excited only at points with density n(**r**) such that the condensate dispersion relation $\omega_{res}(q) = \sqrt{\omega_q^0\left(\omega_q^0 + \frac{2gn}{\hbar}\right)}$ is satisfied, where $\omega_q^0 = \frac{\hbar q^2}{2m} = 2\pi \times 1050\,Hz$. The excitation is phonon-like if $q\xi \ll 1$, and particle-like if $q\xi \gg 1$, where $\xi = \sqrt{1/8\pi na}$ is the local healing length of the gas [8-11]. In our case, $\xi(n_0)^{-1} = 2.64\,\mu m^{-1}$, so that $q\xi(n_0) = 1.60$, and the Bragg excitation is intermediate between the phonon and particle-like regimes. Our lineshape function is expected to be $\omega \times I(\omega)$, where $I(\omega) = \frac{15}{8} \frac{\omega^2 - \omega_q^{0\,2}}{\omega_q^0 (\mu/\hbar)^2} \sqrt{1 - \frac{\omega^2 - \omega_q^{0\,2}}{2\omega_q^0 \mu/\hbar}}$ [11], and where the extra factor of $\omega$ arises from the fact that we measure energy input to the gas, rather than outcoupled atoms. Accounting for our 20% uncertainty in the density as well as the uncertainty in q, we calculate that the first moment of this lineshape function is $\omega_0/2\pi = 1260 \pm 50\ Hz$, in reasonable agreement with our measured value of $1410 \pm 80\ Hz$. This frequency corresponds to resonant excitation at the density $n = 0.57\ n_0$.

The results for non-zero lattice strength show a strong decrease of the resonant frequency with increasing lattice strength. This may be understood qualitatively with Bogoliubov theory [18-21]. To first approximation, the excitations in our experiment can be understood as Bogoliubov sound waves, which have a resonant frequency $\omega_{res} = c_s q$, where $c_s = \sqrt{1/\kappa m^*}$ is the speed of sound, and with $\kappa = [n(\partial \mu/\partial n)]^{-1}$ the compressibility of the gas, and $m^* = (\partial^2 \varepsilon/\partial q^2)^{-1}$ the effective mass, and $\varepsilon(q)$ the energy of the Bloch states of momentum $\hbar q$. As the lattice depth increases the band becomes flatter, and this causes $m^*$ to increase, and the sound speed to decrease. On the other hand, as the lattice depth increases the compressibility increases due to the increased repulsion between the particles in the lattice sites, and this tends to increase the sound speed. However the relative rate of change of the effective mass is much greater than that of the compressibility, so the overall effect is that the sound speed decreases.

In our experiment, there are two effects which cause the peak value of the (unit-cell averaged) density $n_0$ to change with lattice depth $V_0$. One effect is the increasing repulsion between the atoms due to their localization within the lattice sites. The other is an additional contribution to the harmonic trapping force from the optical dipole force of the lattice laser beams, which have a Gaussian intensity profile. These effects can be modeled with a modified Thomas-Fermi approach [22], and for our experiment, we calculate that the peak density changes as shown in Fig. 4(a). For $V_0 = 10\ E_R$, we calculate that the axial and radial trapping frequencies increase to 24.2 Hz and 39.5 Hz, respectively. Going from $V_0 = 0$ to $10\ E_R$, we calculate that the peak density $n_0(V_0)$ decreases from $5.2\times10^{13}$ $cm^{-3}$ to $3.2\times10^{13}$ $cm^{-3}$, corresponding to a decrease in mean site occupancy from 3.7 to 2.3.

We have carried out a numerical calculation of excitation frequencies based on Bogoliubov theory [18-21]. The details of our numerical method can be found in [21]. In the calculation, the trapped Bose gas in an optical lattice is treated as a uniform system. In Fig. 3, we show the result for the density n = 0.57 $n_0(V_0)$. As discussed above, the frequency for this density agrees with the first moment of the theoretical lineshape for zero lattice strength. For non-zero lattice strength the density 0.57 $n_0(V_0)$ should still provide a reasonable definition of "average" density. In Fig. 3, the errors bars on the experimental points account for statistical error of repeated measurements, whereas the error bars on the theoretical points account for the uncertainty in the density and the uncertainty in q. We have confirmed that the calculated frequencies do not depend on the direction of **q** relative to the principal lattice vectors.

The comparison in Fig. 3 shows qualitative agreement with Bogoliubov theory. The main difference is that the experimental values of the resonant frequency decrease more rapidly with lattice strength than the theoretical ones. One factor that may partly account for this is that we move strongly into the phonon-like regime for the higher lattice strengths, which tends to decrease the value of the first moment of the lineshape relative to the resonant frequency in the center of the cloud [10-11]. However, in Fig. 4(b) we show the calculated frequencies for densities $n_0(V_0)$ and the mean density 0.40 $n_0(V_0)$, defined as the number of atoms divided by the volume of the cloud. The calculations show that the observed decrease of the frequency is too large to be accounted for entirely by a modest shift in the first moment of the lineshape function. In Fig. 4(c), we compare the calculated excitation frequency with density $n = 0.57 n_0(V_0)$ with that if the density is held fixed at $n = 0.57 n_0(0) = 3.0\times10^{13}$ $cm^{-3}$. From an optical lattice depth of zero to 10 $E_R$, the calculated excitation frequency decreases by approximately 500 Hz if the density is fixed while the calculated excitation frequency decreases by about 650 Hz if the change in atomic density is taken into account. This shows that the change in density with lattice strength is only a minor factor in the changing resonance frequency in the Bogoliubov approximation.

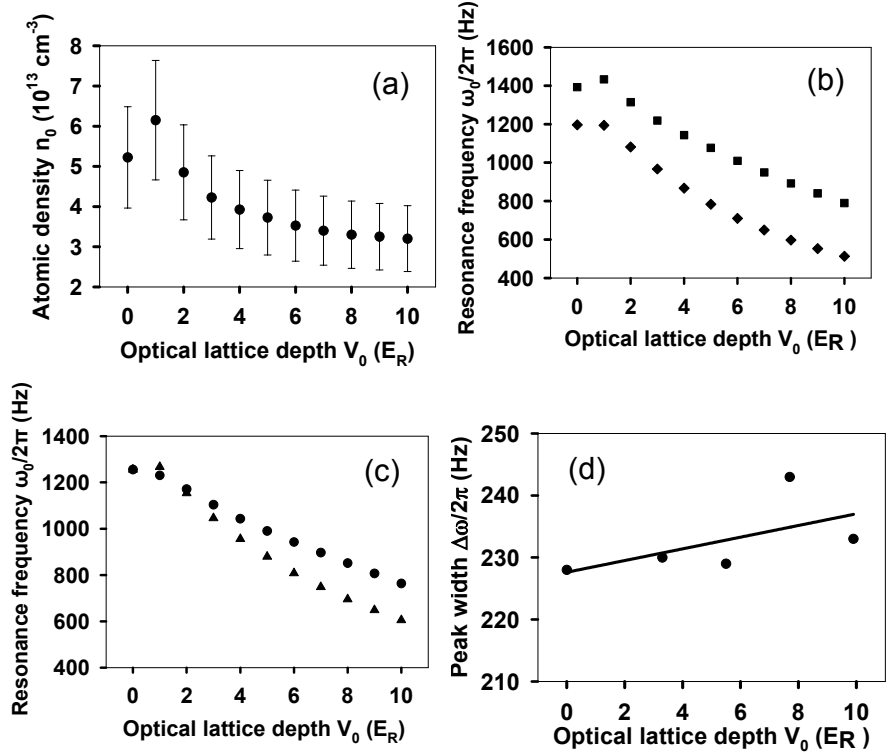

Fig. 4. (a) Peak atomic density $n_0(V_0)$. (b) Calculated resonance frequencies for peak density $n_0(V_0)$ (squares), and for average density $0.40\, n_0(V_0)$ (diamonds). (c) Calculated resonance frequencies for density $n = 0.57\, n_0(V_0)$ (circles), and fixed density $n = 0.57\, n_0(0) = 3.0 \times 10^{13}$ $cm^{-3}$ (triangles). (d) Measured rms peak width versus optical lattice depth. A linear fit to the data is shown.

The modest discrepancy between the theory and experiment may reflect the inaccuracy of our Bogoliubov approximation calculation, which neglects the effects of quantum depletion [16]. Depletion starts to become important when $U > J$, where $U$ and $J$ are the parameters of the Bose-Hubbard Hamiltonian [1,2]

$$H = \sum_i n_i \varepsilon_i - J \sum a_i^+ a_j + U \sum_i \frac{1}{2} n_i (n_i - 1)$$

In this expression the index i labels the lattice sites, $n_i = a_i^+ a_i$ is the atomic number operator for site i, with $a_i^+$ and $a_i$ the raising and lowering operators for site i, and $\varepsilon_i$ is the single particle energy for site i. The second sum is over all pairs of nearest neighbor sites, J is the amplitude for particles to hop between nearest neighbor sites, and U is the contact energy of two particles in the same lattice site. For our parameters, we estimate that $U = J$ for a lattice height of about $V_0 = 3.5\ E_R$. The fact that the slopes of the two curves differ most strongly at about this lattice depth suggests that the onset of depletion may be partly responsible for the difference between the slopes of the theoretical and experimental curves. For a lattice height $V_0 > 13\ E_R$, Bogoliubov theory must fail since the Bose-Hubbard gas becomes an insulator and cannot support long wavelength sound waves. To our knowledge, there has not yet been a calculation of the sound speed of a superfluid Bose-Hubbard gas which takes the effects of depletion into account.

The rms width resonance versus optical lattice depth is given in Fig. 4(d). The width is relatively constant as the lattice strength increases. Note that this may be partly explained by the density dependence of the resonance frequency illustrated in Fig. 4(b), which shows that the spread in the resonance frequencies from the peak to the mean density is roughly constant vs. lattice strength.

In conclusion, we have applied Bragg spectroscopy to measure the linear excitation spectrum of a quantum degenerate gas of bosons in the superfluid regime. The results show that the resonant frequency decreases with increasing lattice strength at a rate that is modestly higher than predicted by a Bogoliubov theory. In the future, Bragg spectroscopy may be useful to determining the excitation spectrum in the insulating phase of the Bose-Hubbard model [12,13] as well as for other quantum gases.

This work is supported by National Science Foundation, the R.A. Welch Foundation, and the Fondren Foundation. Z. X. L and B. W. would like to acknowledge the support from the "BaiRen" program of Chinese Academy of Sciences, the NSF of China (under Grant No. 10674239), and the 973 project of China.

References

[1] D. Jaksch *et al*., Phys. Rev. Lett. **81**, 3108 (1998).

[2] M. P. A. Fisher, P. B. Weichman, G. Grinstein, and D. S. Fisher, Phys. Rev. B **40**, 546 (1989).

[3] M. Greiner *et al.*, Nature (London) **415**, 39 (2002).

[4] D. Jaksch and P. Zoller, Ann. Phys. **315**, 52 (2005).

[5] T. Stöferle *et al.*, Phys. Rev. Lett. **92**, 130403 (2004).

[6] C. Schori *et al.*, Phys. Rev. Lett. **93**, 240402 (2004).

[7] C. Kollath *et al.*, Phys. Rev. Lett. **97**, 050402 (2006).

[8] J. Stenger *et al.*, Phys. Rev. Lett. **82**, 4569 (1999).

[9] D. M. Stamper-Kurn *et al.*, Phys. Rev. Lett. **83**, 2876 (1999).

[10] J. Steinhauer, R. Ozeri, N. Katz, and N. Davidson, Phys. Rev. Lett. **88**, 120407 (2002).

[11] F. Zambelli, L. Pitaevskii, D. M. Stamper-Kurn, and S. Stringari, Phys. Rev. A **61**, 063608 (2000)

[12] D. Van Oosten *et al.*, Phys. Rev. A **71**, 021601 (R) (2005)

[13] A. M. Rey *et al*., Phys. Rev. A **72**, 023407 (2005).

[14] M.-O. Mewes *et al.*, Phys. Rev. Lett. **77**, 416 (1996).

[15] G. Baym and C. J. Pethick, Phys. Rev. Lett. **76**, 6 (1996).

[16] K. Xu *et al.*, Phys. Rev. Lett. **96**, 180405 (2006).

[17] E. G. M. van Kempen, S. J. J. M. F. Kokkelmans, D. J. Heinzen, and B. J. Verhaar, Phys. Rev. Lett. **88**, 093201 (2002).

[18] C. Menotti, M. Krämer, L. Pitaevskii, and S. Stringari, Phys. Rev. A **67**, 053609 (2003).

[19] E. Taylor and E. Zaremba, Phys. Rev. A **68,** 053611 (2003).

[20] D. Boers, C. Weiss, and M. Holthaus, Europhys. Lett. **67**, 887(2004).

[21] Z. X. Liang, X. Dong, Z. D. Zhang, and B. Wu, to be published in Phys. Rev. A

[22] G. K. Campbell *et al.,* Science **313**, 649 (2006); Markus Greiner, Ph. D. Thesis, U. Munich, 2003.